\documentclass[12pt]{article}

\usepackage{amssymb}
\usepackage{array}

\def\0{\over } \def\1{\vec } \def\2{{1\over2}} \def\4{{1\over4}}
\def\5{\bar } 
\def\6{\partial }
\def\7#1{{#1}\llap{/}}
\def\8#1{{\textstyle{#1}}} \def\9#1{{\bf{#1}}}
\def\.{\cdot }
\def\^#1{\widehat{#1}}
 
  \let\g=\gamma \let\d=\delta
   
  \let\l=\lambda \let\m=\mu
    
\let\t=\tau \let\o=\omega 
\let\ph=\varphi

\def\CL{{\cal L}}

\def\({\left(} \def\){\right)} \def\<{\langle } \def\>{\rangle }
\def\[{\left[} \def\]{\right]}  
 
\def\pmbf#1{\setbox0=\hbox{${#1}$}
        \kern-.025em\copy0\kern-\wd0
        \kern.05em\copy0\kern-\wd0
        \kern-.025em\raise.0433em\box0 }

\def\be{\begin{equation}}
\def\ee{\end{equation}}

\newcommand{\bel}[1]{\begin{equation}\label{#1}}
\def\bea{\begin{eqnarray}}
\newcommand{\beal}[1]{\begin{eqnarray}\label{#1}}
\def\eea{\end{eqnarray}}
\def\nn{\nonumber\\ }

\begin{document}

\begin{titlepage}
\renewcommand{\thefootnote}{\alph{footnote}}
\begin{flushright} 
TUW 02-07\\
YITP 02-15
\end{flushright}  
\begin{center}  \vfil 
{\large \bf One-loop surface tensions
of (supersymmetric) kink\\[3pt] domain walls from dimensional regularization}
\vfil 
{\large  
A. Rebhan$^1$\footnote{\footnotesize\tt rebhana@hep.itp.tuwien.ac.at}, 
P. van Nieuwenhuizen$^2$\footnote{\footnotesize\tt vannieu@insti.physics.sunysb.edu} 
and R. Wimmer$^1$\footnote{\footnotesize\tt rwimmer@hep.itp.tuwien.ac.at}
}\\
\end{center}  \medskip \smallskip \qquad \qquad 
{\sl $^1$} \parbox[t]{12cm}{ \sl 
  Institut f\"ur Theoretische Physik, Technische Universit\"at Wien, \\
  Wiedner Hauptstr. 8--10, A-1040 Vienna, Austria\\ } \\
\bigskip \qquad \qquad 
{\sl $^2$} \parbox[t]{12cm}{ \sl 
  C.N.Yang Institute for Theoretical Physics, \\
  SUNY at Stony Brook, Stony Brook, NY 11794-3840, USA\\ } \\
\vfil
\centerline{\large  ABSTRACT}\vspace{.5cm}
We consider domain walls
obtained by embedding the 1+1-dimensional $\phi^4$-kink 
in higher dimensions. We show that a suitably
adapted dimensional regularization method avoids the
intricacies found in other regularization schemes in
both supersymmetric and non-supersymmetric theories.
This method allows us to calculate the one-loop 
quantum mass of kinks and surface tensions of kink domain walls
in a very simple manner, yielding a compact $d$-dimensional
formula which reproduces many of the
previous results in the literature.
Among the new results is the nonvanishing one-loop correction to
the surface tension of a 2+1 dimensional 
$N$=1 supersymmetric kink domain wall with 
chiral domain-wall fermions.

\end{titlepage}

\setcounter{footnote}{0}
\section{Introduction}

One of the simplest situations where one can study quantum
corrections to non-trivial background fields is the calculation
of the quantum mass of 1+1-dimensional solitons
with exactly known fluctuation spectra 
\cite{Raj:Sol,Dashen:1974cj,Coleman:1975bu,Dashen:1975hd,deVega:1976sm,Verwaest:1977tw}. One-loop corrections can be obtained from
computing the difference of the sums (and integrals) of zero-point energies
in the soliton background and in the topologically trivial vacuum.
The regularization of these sums is a surprisingly delicate matter
whose subtleties have been investigated only rather recently,
starting with the observation \cite{Rebhan:1997iv} that for example
a simple energy-momentum cutoff leads to incorrect results
if the same cutoff is used in the topologically distinct sectors.
This has been an actual problem in the calculation of
the quantum mass of
supersymmetric solitons
\cite{Kaul:1983yt,Imbimbo:1984nq,Chatterjee:1984xh,Chatterjee:1985jt,Yamagishi:1984zv}.
On the other hand, the extension of the mode-number cutoff regularization
method introduced by Dashen et al.\cite{Dashen:1974cj}, which
begins by discretizing the problem by means of a finite volume, 
to fermions turns out
to lead to new subtleties concerning the
choice of boundary conditions
which may or may not entail 
a contamination through energies localized at the boundaries 
\cite{Nastase:1998sy,Shifman:1998zy,Goldhaber:2000ab,Wimmer:2001yn}.

However, there do exist
methods which give correct results
that can be formulated a priori in the continuum.
In Ref. \cite{Nastase:1998sy} it has been shown that the derivative
of the quantum kink mass with respect to the mass of elementary scalar
bosons is less sensitive and can be calculated by energy cutoff
regularization, leading to a result for the quantum mass
of susy kinks that agrees with S-matrix factorizations 
\cite{Ahn:1990gn,Ahn:1991uq},
validating also previous results obtained by Schonfeld who considered
mode-number regularization of the kink-antikink system \cite{Schonfeld:1979hg},
and by Refs.~\cite{Cahill:1976im,Boya:1988zh,Casahorran:1989vd} 
using a finite mass formula in terms of only
the discrete modes.
In Refs.~\cite{Farhi:1998vx,Graham:1998kz,Graham:1998qq,Graham:1999pp},
another viable continuum approach was developed that is based on
subtracting successive Born approximations for scattering phase
shifts. Ref.~\cite{Shifman:1998zy} introduced susy-preserving
higher (space) derivative terms in the action and obtained
the correct one-loop results for the energy and the central charge
from simple Feynman graphs. Also heat-kernel and zeta-function
regularization methods have been applied successfully to this
problem \cite{AlonsoIzquierdo:2002eb,Bordag:2002dg}.

In Ref.~\cite{Parnachev:2000fz} it has been shown that
dimensional regularization through embedding kinks as domain
walls in extra dimensions reproduces the known result for
the bosonic kink mass, but it was concluded that this method may
be difficult to generalize.

In the present work, we extend the analysis of Ref.~\cite{Parnachev:2000fz}
and demonstrate that dimensional regularization also allows one
to calculate the surface tensions of kink domain walls
in a way that is far simpler than the methods used previously. 
Moreover, the consideration of domain walls 
gives insight into where precisely naive
cutoff regularization fails, and resolves its ambiguities
by observing that finite ambiguities become divergences
in higher dimensions. Requiring finiteness in $d+1$ dimensions
thus fixes the finite ambiguities in $1+1$ dimensions.
In this way we confirm the recent observation in Ref.~\cite{Litvintsev:2000is}
that the defective energy cutoff method can be repaired
by using smooth cutoffs, or sharp cutoffs as limits of smooth ones.

Through dimensional regularization we derive
a remarkably compact formula for surface tensions
that unifies the diverse results
on kink domain walls in 2+1 and 3+1 dimensions, and yields
a finite result even in 4+1 dimensions. We discuss
the effects of using different renormalization schemes
and confirm (most of the) previous one-loop results in the literature
on kink domain walls in
2+1 and 3+1 dimensions.

We also show that
this way of dimensional regularization works 
for the supersymmetric case by rederiving the quantum mass
of the 1+1 supersymmetric kink, and find a new result
for a 2+1 dimensional supersymmetric kink domain wall
with chiral domain-wall fermions,
which unlike its 3+1 dimensional
analogue has nonzero quantum corrections.

\section{Bosonic kink and domain walls}

\subsection{Bosonic kink and dimensional regularization}

In 1+1 dimensions, 
a real $\ph^4$ theory with spontaneously broken $Z_2$ symmetry
($\ph\to-\ph$)
\bel{Lphi4}
\CL=-\2 (\6_\mu \ph_0)^2-{\l_0\04}(\ph_0^2-\mu_0^2/\l_0)^2
\ee
has topologically non-trivial solutions to the field equations
with finite energy: solitons called ``kinks'',
which interpolate between the two degenerate vacuum states
$\ph=\pm\mu_0/\sqrt{\lambda_0}\equiv \pm v_0$. 
A kink/anti-kink at rest at $x=x_0$ is classically
given by \cite{Raj:Sol}
\bel{Ksol}
\ph_{K,\5K}=\pm 
v_0 
\tanh\(\mu_0(x-x_0)/\sqrt2\).
\ee

Embedding the kink solution in ($d+1$) dimensions instead of (1+1) gives
a domain wall separating the two distinct vacua. This is no longer
a finite-energy solution---its energy is proportional to the transverse
volume $L^{d-1}$, with classical energy density (surface tension)
\bel{M0}
M_0/L^{d-1}=2\sqrt2\mu_0^3/3\lambda_0.
\ee

In $d+1\le 4$ dimensions, (\ref{Lphi4}) is renormalizable
or superrenormalizable, and
upon specifying one's renormalization conditions, quantum
corrections to the energy density should be calculable in perturbation theory
without ambiguity. Some authors are somewhat
cavalier with regard to fixing the meaning of the parameters
of the theory through the renormalization conditions, making
their results basically meaningless: since the lowest order
involves two parameters, any one- or two-loop
result is correct in some renormalization scheme.

In $1+1$ dimensions, 
where kinks correspond to particles with a calculable quantum
mass determined by
the parameters of the Lagrangian, the most frequently used renormalization
scheme consists of demanding that the 
tadpole diagrams cancel
in their entirety, while $\lambda=\lambda_0$ and $\ph=\ph_0$.

Such a renormalization scheme can still be used in $2+1$ dimensions,
whereas in $3+1$ dimensions there is finally the need to
renormalize the coupling constant non-trivially in order to absorb
all one-loop divergences. In the following we shall concentrate
on the particularly natural scheme which fixes
the {\it coupling} constant renormalization such that
in addition to the absence of tadpole diagrams
the renormalized {\it mass} of the elementary scalar be equal to
the pole of its propagator.

Wave-function renormalization $\ph_0=\sqrt{Z}\ph$ is finite to one-loop
order in $3+1$ dimensions and to all orders
in lower dimensions and it is therefore
not mandatory for the
one-loop corrections to the energies of kinks and
kink domain walls.
For simplicity we choose $Z=1$ for now,
postponing the discussion of schemes with nontrivial $Z$ to sect.~\ref{sectZ}.

With $\lambda_0=Z_\lambda \lambda=\lambda+\delta\lambda$,
$v^2_0\equiv \mu^2_0/\lambda_0=Z_{v^2} v^2=v^2+\delta v^2$, the
renormalized Lagrangian for elementary excitations $\eta$ around $\ph=v$
then reads
\beal{Lphi4ren}
\CL&=&-\2 (\6_\mu\eta)^2-\mu^2\eta^2-\mu\lambda^{\2}\eta^3-\4\l\eta^4\nn&&
+\mu\lambda^{\2}\delta v^2 \eta
+\2 (\lambda\delta v^2-2 v^2\delta\lambda)\eta^2
-\4\delta\lambda(\eta^4+4v\eta^3) 
\nn&&-\4(\lambda+\delta\lambda)(\delta v^2)^2
+\2 \delta\lambda \delta v^2(\eta^2+2\eta v),
\eea
which shows that 
the renormalized mass of the elementary boson at tree-graph level
is $m^2=2\mu^2$.

The requirement that tadpole graphs are completely cancelled
by the counterterm proportional to $\eta$ fixes $\delta v^2$
at one-loop level 
\bel{dv2}
\delta v^2=
-3i \hbar\int{dk_0d^{d}k\0(2\pi)^{d+1}}{1\0k^2+m^2-i\varepsilon}=
3\hbar \int{d^{d}k\0(2\pi)^d}{1\02[\vec k^2+m^2]^{1/2}}.
\ee
Using dimensional regularization, where for Euclidean momenta $k_{\mathrm E}$
\be
\int d^{2\nu}k_{\mathrm E}(k^2_{\mathrm E}+M^2)^{-\alpha}=
\pi^\nu (M^2)^{\nu-\alpha}
\Gamma(\alpha-\nu)/\Gamma(\alpha),
\ee
and writing $d=1+s$ so that
$s$ denotes the number of spatial dimensions orthogonal to the
kink axis, we have (setting $\hbar=1$ henceforth)
\bel{dv22}
\delta v^2={3\02\pi} 
(1+s) {\Gamma({-1-s\02})\0\Gamma(-\2)(4\pi)^{s\02}}
\int_0^\infty dk (k^2+m^2)^{s-1\02},
\ee
which is written in a form that will turn out to be convenient
shortly.

Calculating the one-loop correction to the pole mass of the elementary bosons
involves local sea-gull diagrams that are exactly cancelled by $\delta v^2$
and a non-local diagram with 3-vertices. According to (\ref{Lphi4ren})
the renormalized mass $m$ will be equal to the pole mass, if
the latter diagram evaluated on-shell is cancelled by the
counterterm $\propto \delta\lambda \eta^2$. This determines $\delta\lambda$
as
\beal{deltalambda}
\delta\lambda&=&9\lambda^2 {m^{s-2}\0(4\pi)^{s+2\02}} 
\textstyle{\Gamma({2-s\02})}
\int_0^1 dx [1-x(1-x)]^{s-2\02} \nn
&=&9\lambda^2 {m^{s-2}\0(4\pi)^{s+2\02}}\textstyle{\Gamma({2-s\02})
\left(3\04\right)^{s-2\02} {}_2F_1({2-s\02},{1\02};{3\02};-{1\03})},
\eea
where we used ${}_2F_1(a,b;c;z)={\Gamma(c)\0\Gamma(b)\Gamma(c-b)}
\int_0^1 t^{b-1}(1-t)^{c-b-1}(1-zt)^{-a}dt$.
For $s\to2$, i.e. when considering the (3+1)-dimensional theory,
$\delta\lambda$ contains a divergence. For $s<2$, as we
have remarked, the
choice $\delta\lambda=0$ is also a possible renormalization scheme,
and we shall consider it, too, when applicable.

In 1+1 dimensions, the one-loop quantum corrections to the mass
of a kink are determined by the functional determinant of
the differential operator describing fluctuations around the
classical solution (\ref{Ksol}) compared to that of the trivial
vacuum, leading formally to a sum over zero-point energies
which contribute according to
\bel{Msums}
M^{(1)}=M_0+{\hbar\02}\( \sum \o - \sum \o' \) + O(\l)
\ee
where $\o$ and $\o'$ are the eigenfrequencies of fluctuations around
a kink and the vacuum, respectively.
The individual sums as well as their difference are ultraviolet
divergent. The latter divergence is removed by the counter\-terms
obtained by rewriting the bare kink mass $M_0$ in terms of
renormalized parameters
\be\label{M0dM}
M_0 = {2 \sqrt2\03} (v_0^2)^{3/2} \lambda_0^{1/2}
= {m_0^3\0 3\lambda_0} = {m^3\03\lambda} + \delta M
\ee
with 
\be\label{dvlM}
\d M= m \delta v^2 + {m^3\06\lambda^2} \delta\lambda \equiv \delta_v M+
\delta_\lambda M
\ee
or, equivalently, by evaluating the counterterms to the
potential as given by (\ref{Lphi4ren}) in the kink background
\be\label{dvlMint}
\d M=
-{\lambda\02}\,\delta v^2\int_{-\infty}^\infty\[\ph_K^2(x)-{m^2\02\l}\]dx
+{\delta\lambda\04} \int_{-\infty}^\infty\[\ph_K^2(x)-{m^2\02\l}\]^2 dx .
\ee

However, as reviewed in the introduction, the regularization of
the sums over zero-point energies is a highly delicate matter,
and for instance a simple cutoff regularization fails \cite{Rebhan:1997iv}.
Using the same sharp cutoff in energy or, equivalently, momentum
in both the trivial and soliton sector, gives a finite result
where the cutoff can be removed, but this differs from other
regularization procedures by a finite amount. In fact, it has
been shown that cutoff regularization can be repaired by
using smooth cutoffs \cite{Litvintsev:2000is} which are
in fact also required in the calculation of Casimir energies
in order that sums over zero-point energies there can be
evaluated by means of the Euler-McLaurin formula \cite{ItzZ:QFT}.
The limit of a sharp cutoff differs from a straightforward
sharp cutoff by a delta-function peak
in the spectral density at the integration boundary which must not be omitted.
A completely different procedure using sharp cutoffs which
depend on the coordinate $x$ has
recently been proposed in Ref.~\cite{Goldhaber:2001rp} and
independently in Ref.~\cite{Wimmer:2001yn}.  
This ``local mode regularization'' has been used
in Ref.~\cite{Goldhaber:2001rp} to calculate the local distribution
of the quantum energies of 1+1 dimensional solitons.

In the following, we shall however employ dimensional regularization,
which has been shown in Ref.~\cite{Parnachev:2000fz} to
reproduce correctly the quantum mass of the bosonic 1+1 dimensional kink,
and also consider the higher-dimensional kink domain walls.%
\footnote{Dimensional regularization adapted to domain wall
configurations has in fact been discussed already long ago
in Ref.~\cite{Bollini:1984jm}, however without giving
concrete results for the surface tension.}
By analytic continuation of the number $s$ of extra transverse
dimensions of a kink domain wall, no further regularization is
needed. In the vacuum this is indeed consistent with
standard (isotropic) dimensional regularization over $s+1$
spatial dimensions, as its formulae
continue to apply if one first integrates over a subset of
dimensions.

Denoting the momenta pertaining to the $s$ transverse dimensions
by $\ell$ and reserving $k$ for the momentum along the
kink, i.e. perpendicular to the kink domain wall,
the energy of the latter per transverse volume $L^s$ follows
from (\ref{Msums})
\beal{M1kdw}
{M^{(1)}\0L^s}&=&{m^3\03\lambda} + \2 \sum_B \int_{-\infty}^\infty 
{d^s\ell\0(2\pi)^s}
\sqrt{\omega_B^2+\ell^2}\nn&& + \2 \int_{-\infty}^\infty 
 {dk\,d^s\ell \0 (2\pi)^{s+1}}
\sqrt{k^2+\ell^2+m^2}\,\delta_K'(k)+\delta M
\eea
where the discrete sum is over the normalizable states $B$ of the
1+1-dimensional kink with
energy $\omega_B$, 
and the integral is over the continuum part of the
spectrum.

The spectrum of fluctuations for the 1+1-dimensional kink
is known exactly \cite{Raj:Sol}. It consists of a zero-mode,
a bound state with energy $\omega_B^2/m^2=3/4$, and scattering
states in a reflectionless potential for which
the phase shift $\delta_K(k)=-2\arctan(3mk/(m^2-2k^2))$ in the kink background
provides the difference in the spectral density between kink and trivial
vacuum
\be\label{deltaKp}
\int_{-\infty}^\infty dx\, (|\phi_k(x)|^2-1) = 
\delta_K'(k)=-{3m\02}{2k^2+m^2\0(k^2+m^2)(k^2+m^2/4)}.
\ee

The zero mode ($\omega_B=0$), which trivially does not contribute
to the mass of a kink because of its vanishing energy,
corresponds to a massless mode with energy $\sqrt{\ell^2}$ for $s\not=0$,
but does also not contribute to the
energy densities of kink domain walls in dimensional regularization,
because in the latter integrals without a mass scale vanish.
However, it does contribute in cut-off regularization, as
we shall discuss further below.

The leading divergence in the last integral of (\ref{M1kdw})
matches the divergence in $\delta_v M$ and can be combined
with it using (\ref{dv22}) to give (with $x\equiv k/m$)
\beal{M1kdw2}
{M^{(1)}\0L^s}&=&{m^3\03\lambda} + {\Gamma({-1-s\02}) m^{s+1}\0
\Gamma(-\2)(4\pi)^{s\02}} \biggl\{ \2 \left(3\04\right)^{s+1\02} \nn
&&\qquad\quad
+{3\04\pi} \int_{-\infty}^\infty dx (x^2+1)^{s-1\02}
\left[ {-1\04x^2+1}+s \right] \biggr\} + \delta_\lambda M.
\eea
Here the first term inside the braces is the contribution from
the bound state with nonzero energy.

In the limit $s\to0$, which corresponds to the 1+1 dimensional kink,
where one may renormalize ``minimally'' by putting $\delta_\lambda M=0$,
one obtains
\bel{M1DHN}
\Delta M^{(1)}_{s=0}\equiv M^{(1)}_{s=0}-
{m^3\03\lambda} = {m\04\sqrt3} - {3m\02\pi},
\ee
reproducing the well-known DHN result \cite{Dashen:1974cj}.
It is interesting to note that it is the last term in (\ref{M1DHN})
that would be missed in a sharp-cutoff calculation (see Ref.
\cite{Rebhan:1997iv}) and that it
now arises from the last term in the square brackets of (\ref{M1kdw2}).
The latter arises because the counterterm due to $\delta v^2$
does no longer match all of the divergences of the integral
involving $\delta_K'$ for $s>0$, but dimensional regularization
gives a finite result as $s\to0$.

In energy cutoff regularization this term can be recovered by
implementing the cutoff as $\delta(k) \to \delta(k)\theta(\Lambda-k)$
which gives a Dirac-delta in the spectral density by differentiating $\theta$
\cite{Litvintsev:2000is}
and a finite contribution because the scattering phase
$\delta(k)$ decays only like
$1/k$ at large momenta.
The need for such subtle corrections 
is nicely avoided by dimensional regularization:
for sufficiently negative transverse dimensionality $s$ 
the ultraviolet behaviour of
the scattering phases in the longitudinal direction is made harmless.

For $s=1,2,3$, the integral in
(\ref{M1kdw2}) is divergent and gives poles in dimensional
regularization, but as the final results will show, these divergences
are cancelled by the other terms in (\ref{M1kdw2}): for
$s=1,3$, they come from the bound state contribution, whereas for $s=2$,
they are provided by $\delta_\lambda M$.

However, naive cutoff regularization would give rise to problems
which in fact point to the necessity of its modification as
in Ref.~\cite{Litvintsev:2000is}. In contrast to dimensional
regularization, cutoff regularization leads to singularities
for linear and quadratic divergences. Let us consider as an
example the 2+1 case, i.e. $s=1$.
Using a sharp cutoff in the $k$-integral of (\ref{M1kdw}) and 
$\delta M=\delta_v M$,
one can combine these integrals yielding
\bea
&&{M^{(1)}_{s=1}\0L}={m^3\03\lambda} + \int_{-\infty}^\infty
{d\ell\02\pi}\biggl\{
\2\sqrt{\ell^2}+\2\sqrt{\ell^2+3m^2/4}\nn&&
-{1\0\pi}\left[\sqrt{\ell^2}\arctan\sqrt{\ell^2/m^2}+
\sqrt{\ell^2+3m^2/4}\arctan\sqrt{3+4\ell^2/m^2}\right]\biggr\}.
\eea
In this expression, the quadratic divergences cancel (for which it
is necessary that the kink zero-mode is not omitted!), but
because $\arctan(x)=\pi/2-1/x+O(1/x^2)$ for large $x$
the terms in the square bracket also contain linear divergences
that do not cancel. However, if the $k$-integral in (\ref{M1kdw})
is evaluated with a cutoff that is obtained from a smooth
cutoff through a limiting procedure, the Dirac-delta peak
in the spectral density \cite{Litvintsev:2000is} contributes
the additional term 
\be
\lim_{\Lambda_k\to\infty}\int_{-\infty}^\infty
{d\ell\02\pi}{-3m\sqrt{\Lambda_k^2+\ell^2
+m^2}\02\pi\Lambda_k}
\ee
where we have used $\delta(\Lambda_k)\sim 3m/\Lambda_k$.
This renders the complete result finite, and equal to that
obtained in dimensional regularization.

Our study of domain walls thus resolves the ambiguities previously found
in the calculation of the kink mass. Finite ambiguities
in 1+1 dimensions become divergences in $d+1$ dimensions with
$d>1$. Requiring finiteness in $d+1$ dimensions fixes the
finite ambiguities in 1+1 dimensions.

\subsection{Surface tension of bosonic kink domain walls}

For $d>1$, it is straightforward to extract the finite answers for
the one-loop surface tensions of the bosonic kink domain walls by
expanding $s$ around integer values, which leads to elementary
integrals. But instead of giving these individual results, some of
which have been obtained previously, we
shall aim at covering them all together.

\subsubsection{Renormalization schemes with $Z=1$}

First we shall consider renormalization schemes where the wave-function
renormalization constant is kept at $Z=1$ so that $\ph=\ph_0$,
which is a valid and convenient choice at all loop orders for
$s<2$ and to one-loop order for $s=2$.

For general non-integer $s$, the integral in (\ref{M1kdw2}) can be
expressed in terms of the same hypergeometric function
that appeared in the counterterm $\delta\lambda$, eq.~(\ref{deltalambda}),
which was chosen so as to let $m$ coincide with the physical
pole mass of the elementary scalar bosons.\footnote{Using
for example formula (3.259.3) of Ref.~\cite{GraR:T} together
with the linear transformation formulas (9.131), (9.132).}
This leads to the following
remarkably compact formula for the energy densities
of $s$-dimensional bosonic kink domain walls
\bel{MF}
{\Delta M^{(1)}_{\rm OS}\0L^s}=
{m^{s+1}\0(4\pi)^{s+2\02}}{2\Gamma({2-s\02})\0s+1}
\left\{(s+2)\left(3\04\right)^{s\02}
{}_2F_1( {2-s\02}, \2;{3\02};-{1\03})-3 \right\},
\ee
where $m$ is the physical (pole) mass of the elementary scalar, and
the term proportional to $-3$ is produced by the term proportional
to $s$ in (\ref{M1kdw2}).
This is a finite expression for $-1<s<4$.
(The more minimal renormalization scheme where $Z_\lambda=1$,
which is possible for $s<2$ only, is obtained by replacing $(2+s)$
in the first term by 1.)

For the integer values of $s$ of physical interest, the hypergeometric
function in (\ref{MF}) can be reduced to elementary functions
given in Table \ref{tabF}.

\begin{table}
\renewcommand{\arraystretch}{1.5}
\begin{center}
\begin{tabular}{c|l}
\hline
$s$ & ${}_2F_1( {2-s\02}, \2;{3\02};-\kappa)$ \\
\hline
\hline
0 & $\arctan(\sqrt\kappa)/\sqrt\kappa$ \\
1 & ${\rm Arsinh}(\sqrt\kappa)/\sqrt\kappa$ \\
2 & 1 \\
3 & $\2\left[ \sqrt{1+\kappa}+\, 
{\rm Arsinh}(\sqrt\kappa)/\sqrt\kappa\right]$\\
\hline
\end{tabular}
\end{center}
\caption{\label{tabF}\it Special cases of ${}_2F_1$ in (\ref{MF}) for
the values $s$ of physical interest.}
\end{table}

\begin{table}[t]
\renewcommand{\arraystretch}{1.5}
\bigskip
\begin{center}
\begin{tabular}{c|l|l}
\hline
s & ${\Delta M^{(1)}\0L^{s}}/m^{s+1}$ (OS) 
& ${\Delta M^{(1)}\0L^{s}}/m^{s+1}$ (MR)\\
\hline
\hline
0 & ${1\02\sqrt3}-{3\02\pi} \approx -0.189 $ 
& ${1\04\sqrt3}-{3\02\pi}\approx -0.333$ \\
1 & ${3\032\pi}\left( 3\ln3-4 \right)\approx -0.0210$ 
& ${3\032\pi}\left( \ln3-4 \right)\approx -0.0866 $ \\
2 & ${3\016\pi^2}-{1\08\pi\sqrt3}\approx -0.00397$ 
& - \\
3 & $
{9(4-5\ln3)\0(32\pi)^2}\approx -0.00133$
& - \\
\hline
\end{tabular}
\end{center}
\caption{\label{tabM}\it One-loop contributions to 
the quantum mass of the bosonic kink ($s=0$)
and to the surface tension of $s$-dimensional domain walls for the
on-shell (OS), 
and minimal renormalization (MR) schemes, both
with wave-function renormalization $Z=1$.}
\end{table}

\begin{table}[t]
\renewcommand{\arraystretch}{1.5}
\bigskip
\begin{center}
\begin{tabular}{c|l|l}
\hline
s & ${\Delta M^{(1)}\0L^{s}}/m^{s+1}$ (OSR) 
& ${\Delta M^{(1)}\0L^{s}}/m^{s+1}$ (ZM) \\
\hline
\hline
0 & ${2\03\sqrt3}-{2\0\pi} \approx -0.252 $ 
&${1\04\sqrt3}-{19\016\pi}\approx -0.234 $ 
\\
1 & ${5\032\pi}\left( 3\ln3-4 \right)\approx -0.0350$ 
&${3\032\pi}\left( \ln3-{13\06} \right)\approx -0.0319$ 
\\
2 & ${3\08\pi^2}-{1\04\pi\sqrt3}\approx -0.00795$ 
& $-{1\064\pi^2}-{1\032\pi\sqrt3}\approx -0.00733$ 
\\
3 & ${21(4-5\ln3)\0(32\pi)^2}\approx -0.00310$
&  ${-20-9\ln3\0(32\pi)^2}\approx -0.00296$ 
\\
\hline
\end{tabular}
\end{center}
\caption{\label{tabMZ}\it One-loop contributions to 
the quantum mass of the bosonic kink ($s=0$)
and to the surface tension of $s$-dimensional domain walls for the
on-shell scheme with normalized residue (OSR) and
the zero-momentum (ZM) scheme.}
\end{table}

In the 3+1 dimensional case, one has ${}_2F_1(0,\ldots)\equiv 1$,
giving a zero for the content of the braces in eq.~(\ref{MF}), but multiplying
a pole of the Gamma function. Here one has to expand around
$s=2$, for which one needs the following, easily derivable
relation
\bea
&&\lim_{\epsilon\to0} {\Gamma(\epsilon)} \left[ 
{ {}_2F_1(\epsilon, \2;{3\02};-\kappa)}
-1 \right]\nn&=& \sum_{n=1}^\infty {(-\kappa)^n\0n(2n+1)}
= - \ln(1+\kappa)-{2\0\sqrt\kappa}\arctan(\sqrt\kappa)+2\,.
\eea

The numerical results for $s=0,1,2,3$ following from (\ref{MF})
are given in Table \ref{tabM}
for both the physical on-shell renormalization scheme (OS) and,
where applicable, the minimal one with $\delta\lambda=0$ (MR).

\subsubsection{Renormalization schemes with nontrivial $Z$}
\label{sectZ}

Because the kink represents a stationary point of the action, a
nontrivial wave-function renormalization $\ph_0=\sqrt{Z}\ph$
does not introduce additional terms in (\ref{dvlMint}), and
correspondingly leaves the form of the counterterms in
(\ref{M0dM}) and (\ref{dvlM}) unchanged (see the remarks
at the end of section 5.4 in \cite{Raj:Sol}). It does, however,
modify the values of $\delta v^2$ and $\delta\lambda$
in these expressions, and thus changes the numerical result
for $\Delta M^{(1)}$.


In the OS scheme with a nontrivial $Z=1+\delta_Z$, the equation
defining $\delta v^2$ is obtained by replacing
$\delta v^2\to \delta v^2 - v^2 \delta_Z$ in 
the left-hand side of (\ref{dv2})
and the equation defining $\delta\lambda$ by the
substitution $\delta\lambda\to \delta\lambda+\lambda\delta_Z$
in the left-hand side of (\ref{deltalambda})\footnote{In the
latter case there
is a contribution proportional to $\delta_Z$ from the kinetic term,
while the seagull graph now cancels against 
a counterterm with $\delta v^2-v^2\delta_Z$ instead
of a counterterm with only $\delta v^2$.}. 
For any $\delta_Z$ these replacements
in the OS scheme preserve
the relation $\lambda=m^2/(2v^2)$, but with the definition of $m$
fixed, that changes the coupling appearing in the classical
expression $M_{cl.}=m^3/(3\lambda)$ according to
$\lambda=\lambda|_{Z=1}(1-\delta_Z)$. The extra contribution to 
$\Delta M^{(1)}$ is therefore simply $+M_{cl.} \delta_Z$.

A natural refinement of the OS scheme, where $m$ is given by
the physical (pole) mass, is to require that the residue of
this pole be unity. This leads to
\bea
&&\delta_Z=-9\lambda {m^{s-2}\0(4\pi)^{s+2\02}}
\textstyle{\Gamma({4-s\02})\int_0^1dx\,x(1-x)[1-x(1-x)]^{s-4\02}}\nn
&&=9\lambda {m^{s-2}\0(4\pi)^{s+2\02}} \textstyle{\Gamma({4-s\02}) 
({3\04})^{s-2\02}
\biggl[ {}_2F_1({2-s\02},{1\02};{3\02};-{1\03})
-{4\03}\,{}_2F_1({4-s\02},{1\02};{3\02};-{1\03}) \biggr].}\qquad
\eea

Curiously enough, with the help of Gauss' recursion relations \cite{GraR:T}
the particular combination of hypergeometric
functions in this expression can be recast in a form
proportional to (\ref{MF}),
\be
\delta_Z= 2\lambda{m^{s-2}\0(4\pi)^{s+2\02}}\Gamma({2-s\02})
\left\{(s+2)\left(3\04\right)^{s\02}
{}_2F_1( {2-s\02}, \2;{3\02};-{1\03})-3 \right\}.
\ee

The energy densities of kink domain walls in an on-shell renormalization
scheme with physical pole mass and unit residue (OSR) is thus
given by the simple conversion formula
\be\label{MFOSR}
{\Delta M^{(1)}_{\rm OSR}\0L^s}={\Delta M^{(1)}_{\rm OS}\0L^s}+
{m^3\03\lambda}\delta_Z={s+4\03}{\Delta M^{(1)}_{\rm OS}\0L^s}
\ee
and the particular results for the values $s$ of interest
are listed in Table \ref{tabMZ}.

For the sake of comparison with previous results in
the literature, Table \ref{tabMZ}
also includes another widely used 
renormalization scheme \cite{Luscher:1988ek}, 
where the
mass is renormalized at zero momentum (ZM) according to
$m^2_{ZM}=\Gamma^{(1)}(0)$ 
with $\Gamma^{(1)}(k^2)$ the inverse propagator to one-loop order
and $\delta_Z$ is chosen such that 
$[\partial\Gamma^{(1)}/\partial{k^2}](0)=1$.
In this scheme, formula~(\ref{MF}) gets replaced by
\beal{MFZM}
{\Delta M^{(1)}_{\rm ZM}\0L^s}&=&
{m^{s+1}\0(4\pi)^{s+2\02}}{2\Gamma({2-s\02})\0s+1}
\biggl\{\left(3\04\right)^{s\02}
{}_2F_1( {2-s\02}, \2;{3\02};-{1\03}) 
\nn
&&\qquad\qquad +{3\04}(s-3)-{1\016}(s+1)(2-s) 
\biggr\},
\eea
where the very last term within the braces is the contribution of $\delta_Z$.

The surface tension of $\ph^4$ domain walls has been calculated
in the ZM scheme
to one-loop order in 3+1 dimensions in Ref.~\cite{Munster:1989we}
by considering the energy splitting of the two lowest states
in a finite volume using zeta-function techniques,
and our result completely agrees with that.
Our result
is also consistent
with the older Ref.~\cite{Brezin:1984}
using $\epsilon$-expansion (in the limit $\epsilon\to0$), which
employed yet another renormalization scheme that is
closer (but not identical) to an $\overline{\mbox{MS}}$-scheme.
We do not, however, agree with the ZM-scheme result reported in
Ref.~\cite{deCarvalho:1986sc} nor with its correction
in Ref.~\cite{deCarvalho:2001da}
\footnote{The latter reports the same result
as that contained in Ref.~\cite{Brezin:1984} (for $\epsilon\to0$), 
while formulating different
renormalization conditions amounting to the ZM scheme at one-loop order.}.

In 2+1 dimensions, the surface
tension of the kink domain wall has been
calculated in Ref.~\cite{Munster:1990yg}, and 
in Ref.~\cite{Hoppe:1998ey} to two-loop order
in the ZM scheme. Our one-loop ZM result reproduces that
given in Ref.~\cite{Hoppe:1998ey}, while the one-loop
result of Ref.~\cite{Munster:1990yg} cannot be directly compared
with ours as it re-expresses the ZM result in terms of
the physical pole mass without using the coupling of
either our OS or OSR scheme.
We also agree with the
most recent work
\cite{Graham:2001kz}, where the 2+1 dimensional kink domain
wall energy density was calculated using the Born approximation
methodology of Refs.~\cite{Farhi:1998vx,Graham:2001dy}
in the MR scheme.
Compared to Ref.~\cite{Graham:2001kz}, the present calculation
in dimensional regularization turns out to be considerably simpler
and more straightforward,
as the former has to exert some care in identifying ``half-bound''
states and to employ certain non-trivial sum rules for phase shifts.
On the other hand, the methods of \cite{Farhi:1998vx,Graham:2001dy}
will be useful also in cases where one can determine phase shifts
only numerically.

Comparing finally the size of the one-loop corrections in the four
different renormalization schemes considered in Tables~\ref{tabM}
and \ref{tabMZ}, 
one notices that the corrections
are largest in the MR scheme and significantly smaller in
the other schemes, with the ZM and OSR results being rather
close, but with noticeable differences.

These issues are of relevance in practical applications,
and, indeed, the surface tension of the $\ph^4$ kink
domain wall can be related to universal quantities that can
be investigated by lattice simulations of the
Ising model
and experimentally in binary 
mixtures \cite{Hoppe:1998ey}. 
In a comparison of the field-theoretic results
with lattice studies, the different definitions of
mass in the OS and in the ZM scheme correspond to
the true (exponential) correlation length and
to the second moment of the correlation function,
respectively, both of which can
be found in the literature
(see e.g.~\cite{Hasenbusch:1997} and references therein).

Of perhaps mere academic interest is the case of kink domain walls
in 5 dimensions ($s=3$) where our formulae still give finite results.
In 5 dimensions, $\ph^4$ theory is of course no longer renormalizable,
though it may still be of interest as an effective theory.

\section{Supersymmetric kink and domain walls}

\subsection{The susy kink and domain string} 

In 1+1 and 2+1 dimensions ($s=0$ and $s=1$), the model (\ref{Lphi4}) has
the supersymmetric extension
\cite{DiVecchia:1977bs,Hruby:1977nc}
\bel{Lss}
\CL=-\2\[ (\6_\m\ph)^2+U(\ph)^2+\5\psi\g^\m\6_\m\psi+U'(\ph)\5\psi\psi \]
\ee
where $\psi$ is a Majorana spinor, $\5\psi=\psi^{\mathrm T} C$ and
\be\label{Uphi}
U(\ph)=\sqrt{\l_0\02}\(\ph^2-v_0^2\),\qquad v_0^2\equiv \mu_0^2/\l_0.
\ee
(In 1+1 dimensions, $U\propto \sin(\sqrt\gamma\ph/2)$ gives
the sine-Gordon model, which is however not renormalizable in
2+1 dimensions.)

Imbedding the susy kink in 2+1 dimensions gives a 
domain wall centered about a one-dimensional string on which
the fermion mass vanishes (since $U'(\ph_K)\propto \ph_K$ vanishes
at the center of the kink). In the following we shall succinctly
refer to this particular domain wall as ``domain string'', postponing
a brief discussion of higher-dimensional kink domain walls to the next
subsection.

Going from 1+1 to 2+1 dimensions, the discrete symmetry content
of (\ref{Lss}) in fact changes. In 1+1 dimensions, (\ref{Lss})
has the $Z_2$ symmetry $\ph\to-\ph$, $\psi\to \gamma^5 \psi$
with $\gamma^5=\gamma^0\gamma^1$.
In 2+1 dimensions, on the other hand, $\gamma^5=\gamma^0\gamma^1\gamma^2
\propto \pm {\bf 1}$, and the sign of the fermion mass term can no longer be
reversed by $\psi\to \gamma^5 \psi$. By the same token,
(\ref{Lss}) breaks parity, because a sign change of one of
the spatial $\gamma$ matrices cannot be effected by an
equivalence transformation, but leads to the other of the two
inequivalent irreducible representations
of a Clifford algebra in odd space-time dimensions.

In what follows we shall consider the quantum corrections to
both, the mass of the susy kink and the tension of the
domain string, together. In both cases we shall
continue to use a renormalization scheme where we put
$Z_\ph=1=Z_\psi$ at one-loop order. For this reason we have already
dropped a subscript 0 for the unrenormalized fields in (\ref{Lss}).
We shall however consider the possibility of (finite)
coupling constant renormalization, again by requiring
that the renormalized mass of elementary scalars and fermions
be given by the physical pole mass, together with the
requirement of vanishing tadpoles, which fixes $\delta v^2$.

Inclusion of the fermionic tadpole loop replaces $3$ by $(3-2)$
in (\ref{dv2}) so that compared to the bosonic result we have
$$\delta v^2|_{\rm susy}\equiv \delta \tilde v^2={1\03}\delta v^2|_{\rm bos.}$$
(When useful we distinguish quantities in the susy case by twiddles.)

In the OS scheme, the supersymmetric version of (\ref{deltalambda})
is obtained by the replacement\footnote{The counterterm
$\2 \lambda \delta v^2 \eta^2$ induced by the tadpole with a fermionic loop
cancels only those contributions to the bosonic selfenergy due to a
fermionic loop which contain one propagator.
The remaining contributions
have two propagators and are proportional to the bosonic contribution to
the selfenergy.}
$$9m^2\to  9m^2-2(2m^2+\2 q^2)|_{q^2=-m^2}=6m^2,$$
and thus
$$\delta \tilde\lambda={2\03}\delta\lambda|_{\rm bos.}.$$

In a Majorana representation of the Dirac matrices 
in terms of the usual Pauli matrices $\tau^k$
with $\g^0=-i\t^2$,
$\g^1=\t^3$, $\gamma^2=\tau^1$ (added for $s=1$), 
and $C=\t^2$ so that $\psi={\psi^+\choose\psi^-}$ with
real $\psi^+(x,t)$ and $\psi^-(x,t)$, the equations for the
bosonic and fermionic normal modes with frequency $\omega$ 
and longitudinal momentum $\ell$ (nonzero only when $s=1$) 
in the kink background $\ph=\ph_K$ read
\bea
&&[-\partial_x^2+U'{}^2+UU'']\eta=(\omega^2-\ell^2)\eta, \\
&&(\partial_x+U')\psi^++i(\omega+\ell)\psi^- = 0, \label{psip}\\
&&(\partial_x-U')\psi^-+i(\omega-\ell)\psi^+ = 0\label{psim}.
\eea
Acting with $(\partial_x-U')$ on (\ref{psip}) and eliminating
$\psi^-$ as well as $\ph'=-U$ shows that $\psi^+$ satisfies
the same equation as the bosonic fluctuation $\eta$.
Compared to $\psi^+$, the component $\psi^-$ has a continuous
spectrum whose modes differ by an additional phase shift
$\theta=-2\arctan(m/k)$ when traversing
the kink from $x_1=-\infty$ to $x_1=+\infty$, which is
determined only by $U'(\ph_K(x_1=\pm\infty))=
\pm m$.
Correspondingly, the difference of the spectral densities of the
$\psi^+$-fluctuations in the kink and in the trivial vacuum
equals that of the $\eta$-fluctuations, given in (\ref{deltaKp}),
whereas that of $\psi^-$-fluctuations
is obtained by replacing $\delta'_K\to \delta'_K+\theta'$.

In the sum over zero-point energies
for the one-loop quantum mass of the kink (when $s=0$),
\bel{Msumbf}
\tilde M=\tilde M_{cl.}+\2\(\sum\o_B-\sum\o_B'\)-\2\(\sum\o_F-\sum\o_F'\)
+\d \tilde M \;,
\ee
one thus finds that the bosonic contributions from
the continuous spectrum are canceled by the fermionic
contributions except for the additonal contribution
involving $\theta'(k)$ in the spectral density of the $\psi^-$ modes. 

The discrete bound states cancel
exactly, apart from the subtlety that the fermionic zero
mode should be counted as half a fermionic mode \cite{Goldhaber:2000ab}.
In strictly 1+1 dimensions, the zero modes do not contribute
simply because they carry zero energy, and for $s>0$, where
they become massless modes, they
do not contribute in dimensional regularization.

In a cutoff regularization in $s=1$, as we already discussed and
shall further discuss below,
they in fact do play a role. Remarkably, the half-counting of
the fermionic zero mode for $s=0$ has an analog for $s=1$ where the
bosonic and fermionic zero modes of the kink 
correspond to massless modes with energy
$|\omega|=|\ell|$. From (\ref{psip}) and (\ref{psim}) one
finds that the fermionic kink-zero mode $\psi^+\propto \ph_K'$,
$\psi^-=0$
is a solution only for $\omega=+\ell$. It therefore cancels
only half of the contributions from the bosonic kink-zero mode
which for $s=1$ have $\omega=\pm\ell$.
For $s=1$ one thus finds that the fermionic zero mode of the
kink corresponds to a chiral (Majorana-Weyl) fermion on
the ($s$=1)-dimensional domain string
\cite{Callan:1985sa,Gibbons:2000hg,Hofmann:2000pf}.%
\footnote{Choosing a different sign for $\gamma_1$
reverses the allowed sign of $\ell$ for these fermionic modes
and thus their chirality (with respect to the domain string
world sheet). This
corresponds to the other, inequivalent representation of the
Clifford algebra in 2+1 dimensions.}

In dimensional regularization, however, the kink zero modes and
their massless counterparts for $s>0$ can
be dropped, and the energy density of the
susy domain wall reads
\bel{M1skdw}
{\tilde M^{(1)}\0L^s}={m^3\03\lambda} - \4 \int {dk\,d^s\ell \0 (2\pi)^{s+1}}
\sqrt{k^2+\ell^2+m^2}\,\theta'(k)+\delta \tilde M,
\ee
where
\be
\theta'(k)={2m\0k^2+m^2}.
\ee

With 
$\delta_v \tilde M = {1\03} \delta_v M $ the logarithmic
divergence in the integral in (\ref{M1skdw}) as $s\to0$
gets cancelled. A naive cut-off regularization at $s=0$ would
actually lead to a total cancellation of the $k$-integral with
the counterterm $\delta_v \tilde M$, giving a vanishing
quantum correction in renormalization schemes with $\lambda=\lambda_0$.
In dimensional regularization there is now however a mismatch
for $s\not=0$ and a finite remainder in the limit $s\to0$
proportional to $s\Gamma(-s/2)$. Including the
optional $\lambda$-renormalization the final result reads
\bel{M1skdw2}
{\tilde M^{(1)}\0L^s}={m^3\03\lambda} 
-{m^{s+1}\0(4\pi)^{s+2\02}}{2\Gamma({2-s\02})\0s+1}
+ \delta_\lambda \tilde M.
\ee

In the minimal renormalization (MR) scheme one has 
$\delta_\lambda \tilde M=0$,
whereas in the more physical OS scheme, where $m$ is the pole
mass of the elementary bosons as well as fermions, one
has $\delta_\lambda \tilde M = {2\03} \delta_\lambda M$, yielding
\bel{MFs}
{\Delta \tilde M^{(1)}\0L^s}=
{m^{s+1}\Gamma({2-s\02})\0(4\pi)^{s+2\02}}
\left\{\left(3\04\right)^{s-2\02}
\!{}_2F_1( {2-s\02}, \2;{3\02};-{1\03})-{2\0s+1} \right\}.
\ee

\begin{table}[t]
\renewcommand{\arraystretch}{1.5}
\bigskip
\begin{center}
\begin{tabular}{c|l|l}
\hline
s & ${\Delta \tilde M^{(1)}\0L^{s}}/m^{s+1}$ (OS) & ${\Delta \tilde M^{(1)}\0L^{s}}/m^{s+1}$ (MR)\\
\hline
\hline
0 & ${1\06\sqrt3}-{1\02\pi} \approx -0.063 $ & $-{1\02\pi} \approx -0.159$\\
1 & ${1\08\pi}(\ln3-1) \approx +0.004$ & $-{1\08\pi} \approx -0.040$\\
\hline
\end{tabular}
\end{center}
\caption{\label{tabMs}\it One-loop contributions to 
the quantum mass of the susy kink ($s=0$)
and to the surface tension of the 
(s=1)-dimensional susy kink domain ``wall'' for the
on-shell (OS) and minimal renormalization (MR) schemes.}
\end{table}

The respective results for the 1+1 dimensional susy kink ($s=0$)
and for the (s=1)-dimensional susy kink domain ``wall'' (domain string)
are given in
Table~\ref{tabMs}. Again we find that there is much faster apparent
convergence in the OS scheme compared to the MR one where only
the tadpoles are subtracted.

In the literature, at least to our knowledge,
only the case of a supersymmetric kink ($s=0$) in the MR scheme\footnote{In
Refs.~\cite{Kaul:1983yt,Rebhan:1997iv} the respective results
have also been expressed in terms of the physical pole mass,
but keeping $\lambda$ as in the MR scheme. Such a renormalization
scheme yields a tadpole contribution proportional to $\delta\lambda$
and should  not be confused with the OS scheme considered here,
where both the mass and the coupling is renormalized such as to
have both vanishing tadpoles and a physical pole mass for the
elementary bosons.}
has been considered and dimensional regularization reproduces
the result obtained before by 
Refs.~\cite{Schonfeld:1979hg,Boya:1988zh,Nastase:1998sy,Graham:1998qq,Shifman:1998zy}. 
However, a (larger) number of papers have missed
the contribution $-m/(2\pi)$ because of the (mostly implicit) use
of the inconsistent energy cutoff scheme
\cite{Kaul:1983yt,Imbimbo:1984nq,Chatterjee:1984xh,Chatterjee:1985jt,Yamagishi:1984zv}
or have obtained different answers because of the use
of boundary conditions that accumulate a finite amount of
energy at the boundaries \cite{Uchiyama:1986gf,Rebhan:1997iv}.
The former result is however now generally accepted and,
in the case of the super-sine-Gordon model (where the
same issues arise with the same results) in agreement with
S-matrix factorization \cite{Ahn:1991uq}.

In Ref.~\cite{Litvintsev:2000is} the correct susy kink mass
has also been obtained by employing a smooth energy (momentum) cutoff,
the necessity of which becomes apparent, as in the purely bosonic
case, by considering the 2+1 dimensional domain wall.
Using a naive cutoff for $s=1$ one finds quadratic divergences
which cancel only upon inclusion of the zero modes (which become massless
modes in 2+1 dimensions). As we have
discussed above, unlike the other bound states, these do not
cancel because the fermionic zero mode becomes a chiral
fermion on the domain-string world-sheet and thus 
cancels only half of the bosonic zero (massless) mode contribution,
yielding
\bea
&&\int_0^\infty {d\ell\02\pi} \biggl\{\2\sqrt{\ell^2}
-\int_{-\Lambda_k}^{\Lambda_k}{dk\02\pi} 
\left[\sqrt{k^2+\ell^2+m^2}{m\0k^2+m^2}-{1\0\sqrt{k^2+\ell^2+m^2}} \right]
\biggr\}\nn
&&
\stackrel{\Lambda_k\to\infty}{\longrightarrow}
\int_0^\infty{d\ell\02\pi} \biggl\{{\ell\02}-{\ell\0\pi}
\arctan{\ell\0m}\biggr\}
\sim\int_0^\infty{d\ell\0\pi}{m\02\pi}
\eea
which is however still linearly divergent. Smoothing out the
cutoff in the $k$-integral does pick an additional (and for
$s=0$ the only) contribution $-m/(2\pi)$, which is now
necessary to have a finite result for $s=1$. This finite
result then reads
\be
{\tilde M^{(1)}_{s=1}\0L}=-{1\0\pi}\int_0^\infty {d\ell\02\pi}
\left(m-\ell\arctan{m\0\ell}\right)=-{m^2\08\pi}
\ee
in agreement with the result obtained above in dimensional regularization.

\subsection{Susy kink domain walls in 3+1 dimensions}

For completeness we shall also briefly discuss kink domain
walls in the 3+1-dimensional
Wess-Zumino-model \cite{Wess:1974kz}. 
In accordance with Ref.~\cite{Dvali:1997bg,Chibisov:1997rc}
we shall demonstrate that in this model there is no nontrivial
quantum correction to the surface tension.

A Wess-Zumino model with a spontaneously broken $Z_2$ symmetry
now requires two real scalar fields to pair up with
the now four-component Majorana spinor. For the classical
Lagrangian we choose
\begin{eqnarray}
  \label{eq:wz-lagrange}
  &&{\cal L}=-\2 (\partial A)^2-\2(\partial B)^2-V(A,B)
  -\2 \bar{\psi}[\not\!\partial + \sqrt{2\lambda}(A+i\gamma_5 B)]\psi 
\nonumber \\
  &&V(A,B)=\frac{\lambda}{4}(A^2-B^2-v^2)^2+\lambda A^2B^2,
\end{eqnarray}
where $A$ is a real scalar with non-vanishing vacuum expectation value,
while $B$ is a real pseudo-scalar without one. For $B\equiv0$
the potential coincides with that of the kink model (\ref{Lphi4}),
and correspondingly a classical domain wall solution is
given by $A_K(x)=\phi_K(x_1)$ and all other fields zero.

As is well known \cite{Iliopoulos:1974zv,Grisaru:1979wc}, 
in the 3+1-dimensional
Wess-Zumino-model there is only one non-trivial renormalization
constant $Z$ for the kinetic term, which implies
$\mu^2=Z \mu_0^2$ and $\lambda=Z^{3/2}\lambda_0$ and thus
a vanishing counter-term $\delta M$ for the kink wall energy density.

The fluctuation equations for $\eta=A-A_K$, $B$, and $\psi$ read
\begin{eqnarray}
  \label{eq:fluct1}
  && \partial^2 \eta -({U'}^2+UU'')\eta=0 \nonumber \\
  && \partial^2 B-(\lambda A_K^2+\mu^2)B=0 \nonumber \\
  && [\not\!\partial + U']\psi=0,
\end{eqnarray}
with $U$ as in (\ref{Uphi}).
$A_K$ satisfies the Bogomolnyi equation 
$A_K'=-U(A_K)$, and the $x$-dependent parts of the $\eta$ and $B$ field 
equations factorize as $-(\partial_x-U')(\partial_x+U')$ 
and $-(\partial_x+U')(\partial_x-U')$, respectively.

Both the $\eta$ and $B$ fluctuation equations involve
reflectionless potentials of the form
\begin{equation}
  \label{eq:reflectionslos}
  -\partial_z^2-\frac{n(n+1)}{\cosh^2z}+n^2,
\end{equation}%
where $z:=\frac{mx}{2},\; m:=\sqrt{2}\mu$.

The kink fluctuation modes $\phi_K(x)$ correspond to $n=2$, 
and the $\eta$ fluctuations are given by the former multiplied
by plane waves with momentum $\vec\ell=(\ell_2,\ell_3)$
in the trivial directions. 
Their spectrum thus consists of one massless mode and one
massive mode localized on the domain wall with
$\omega^2_0(\ell)=\ell^2$ and  $\omega^2_B(\ell)={3\04}m^2+\ell^2$
and delocalized ones with $\omega^2_k(\ell)=k^2+\ell^2+m^2$.

The $x$-dependence of the $B$-fluctuations on the other hand
involves the potential (\ref{eq:reflectionslos}) with $n=1$,
like the fluctuation equations for the sine-Gordon soliton,
but with different energies according to
\begin{equation}
  \label{eq:sg}
   \left(-\partial_z^2-\frac{2}{\cosh^2z}+1\right)s(z)=
[\frac{4}{m^2}(\omega^2-\ell^2)-3]s(z).
\end{equation}
The spectrum of the sine-Gordon system is now shifted by
$\ell^2+{3\04}m^2$ so that the sine-Gordon
zero-mode matches the bound state of the kink, and the continuous part
of the spectrum also coincide. The spectrum of the $B$-fluctuations
thus equals that of the $\eta$-fluctuations apart from
the absence of the massless (zero) mode. The spectral densities
for the delocalized modes are, however, different and the bosonic
contribution to the one-loop surface tension reads
\be
{\Delta^{\rm b}\tilde M^{(1)}\0L^s}=
\2 \int {d^s\ell\0(2\pi)^s} \left(
\omega_0(\ell)+2 \omega_B(\ell)+\int{dk\02\pi} \omega_k(\ell)
[\delta'_K(k)+\delta'_{SG}(k)] \right)
\ee
where $s=2-\epsilon$.

Choosing the Majorana representation for the Dirac matrices
\be
  \label{eq:gammas}
  \gamma^0=\left({0\atop1}{-1\atop0}\right)\ ,
\gamma^1=\left({1\atop0}{0\atop-1}\right),\ 
  \gamma^2=\left({0\atop\tau_1}{\tau_1\atop0}\right),\ 
  \gamma^3=\left({0\atop\tau_3}{\tau_3\atop0}\right), 
\ee
and writing $\psi$ in terms of two 2-component spinors
$e^{-i\omega t+\ell\cdot x_\parallel}
(\psi_A,\psi_B)$, the fermionic fluctuation equation
of (\ref{eq:fluct1}) becomes
\begin{eqnarray}
  \label{eq:fermfluct}
  &&(\partial_x+U')\psi_A + i[\omega+\not\!\ell]\psi_B=0 
\label{ferm1}\\
  &&i[\omega-\not\!\ell]\psi_A+(\partial_x-U')\psi_B=0 
\label{ferm2},
\end{eqnarray}
where $\not\!\ell=\tau_1\ell_2+\tau_3\ell_3$.
Through (\ref{ferm1}), $\psi_B$ can be expressed algebraically
in terms of $\psi_A$, except when $\omega^2=\ell^2$,
and inserting into (\ref{ferm2})
shows that the latter satisfies the same fluctuation equation
as the bosonic fluctuation $\eta$.
Using that $(\partial_x+U')\phi_k = \omega_{Kink} s_k$, one
finds that $\psi_B$ has the same spectrum as the $B$ fluctuations.

For the massless (zero) mode ($\omega_{Kink}=0$) only
$(\partial_x+U')\psi_A=0$ in (\ref{ferm1}) has a normalizable solution, 
which is located at the domain wall. The other equation, 
$(\partial_x-U')\psi_B=0$, has normalizable solutions only
if boundaries for the $x$-direction were introduced, and
would be localized there.

As a result, the fermionic contribution to the one-loop
correction of the domain wall tension becomes identical to
the bosonic one, but with a negative sign,
\be
{\Delta^{\rm f}\tilde M^{(1)}\0L^s}=-{\Delta^{\rm b}\tilde M^{(1)}\0L^s}.
\ee
In perfect agreement with the non-renormalization theorem of
the superpotential (which does not apply at the lower dimensions
considered above), there is no quantum correction to the
classical value of the surface tension of the susy kink
domain wall in 3+1 dimensions.

This cancellation of the quantum corrections can also be linked
to the cancellation of quantum corrections to the $N=2$ susy
kink mass \cite{Nastase:1998sy,Shifman:1998zy}.

Such a cancellation is also to be expected for 4+1 dimensional
supersymmetric theories with domain walls. In contrast to
2+1 dimensions, in 4+1 dimensions there are no Majorana fermions,
so one needs to extend the supersymmetry algebra to involve a
Dirac fermion. From the point of view of the 1+1 dimensional
kink, this will imply $N=4$ supersymmetry. On the then 4-dimensional
domain wall one may have chiral fermions, but as pointed out
in Ref.~\cite{Gibbons:2000hg}, these domain-wall fermions necessarily
come in pairs containing both chiralities.

\section{Conclusion}

In this paper we have shown that dimensional regularization
allows one to compute the one-loop contributions to the quantum
energies of bosonic and supersymmetric kinks and kink domain walls
in a very simple manner. The ambiguities associated with
ultraviolet regularization observed in the 1+1 dimensional
kinks has been shown to be eliminated by considering their
extension to kink domain walls in higher dimensions.

For the bosonic kink domain walls, which are of interest
also in the context of condensed matter physics, we have
derived a compact $d$-dimensional formula, which reproduces
and (mostly) confirms existing results in the literature, and
we have also discussed in detail the dependence on particular 
renormalization schemes.

In the supersymmetric case, we confirmed previous results
in 1+1 and 3+1 dimensions. While in the latter case quantum
corrections to the surface tension vanish, we have obtained
a nontrivial one-loop correction for a 2+1 dimensional $N=1$
susy kink domain wall with chiral domain wall fermions.
The nontrivial quantum corrections to the supersymmetry algebra in
the 1+1 and 2+1 dimensional models will be discussed
in a forthcoming publication.

\subsection*{Acknowledgments}

The authors are much indebted to Gernot M\"unster (University of
M\"unster) for a clarifying correspondence.
P.v.N. thanks the Technical University of Vienna and Wolfgang Kummer for
einen wundervollen Aufenthalt of two months. R.W. has been
supported by the Austrian Science Foundation, project
no. P15449.

\bibliographystyle{h-physrev}
\bibliography{qft,ar,books}

\end{document}